\documentclass{aa}  
\usepackage{graphicx}
\usepackage{eqnarray,amsmath}
\usepackage{mathtools}
\usepackage{multicol}
\usepackage[hypcap]{caption}
\usepackage{txfonts}
\usepackage{subcaption}
\usepackage{arydshln}
\usepackage{hyperref}
\usepackage{xcolor}
\graphicspath{{./figures/}} 
\newcounter{daggerfootnote}

\begin{document} 

\titlerunning{Evolution of massive stars adopting self-consistent winds at SMC metallicity}
\title{Evolution of rotating massive stars adopting a newer, self-consistent wind prescription at SMC metallicity}
\authorrunning{Gormaz-Matamala et al.}
\author{A. C. Gormaz-Matamala\inst{1,2,3}
\and
J. Cuadra\inst{2}
\and
S. Ekström\inst{4}
\and
G. Meynet\inst{4}
\and
M. Curé\inst{5}
\and
K. Belczynski\inst{1}\thanks{Deceased on 13th January 2024.}}
\institute{
Nicolaus Copernicus Astronomical Center, Polish Academy of Sciences, Bartycka 18, 00-716 Warsaw, Poland\\
\email{agormaz@camk.edu.pl}
\and
Departamento de Ciencias, Facultad de Artes Liberales, Universidad Adolfo Ib\'a\~nez, Av. Padre Hurtado 750, Vi\~na del Mar, Chile
\and
Instituto de Astrof\'isica, Facultad de F\'isica, Pontificia Universidad Cat\'olica de Chile, 782-0436 Santiago, Chile
\and
Geneva Observatory, University of Geneva, Maillettes 51, 1290 Sauverny, Switzerland
\and
Instituto de Física y Astronomía, Universidad de Valparaíso. Av. Gran Breta\~na 1111, Casilla 5030, Valpara\'iso, Chile.}

\date{}

\abstract 
{}
{We aim to measure the impact of our mass loss recipe in the evolution of massive stars at the metallicity of the Small Magellanic Cloud (SMC).}
{We use the Geneva-evolution-code to run evolutionary tracks for stellar masses ranging from $20$ to $85$ $M_\odot$ at SMC metallicity ($Z_\text{SMC}=0.002$).
We upgrade the recipe for stellar winds by replacing Vink's formula with our self-consistent m-CAK prescription, which reduces the value of mass-loss rate, $\dot M$ by a factor between $2$ and $6$ depending on the mass range.}
{The impact of our new [weaker] winds is wide, and it can be divided between direct and indirect impact.
For the most massive models ($60$ and $85$ $M_\odot$) with $\dot M\gtrsim2\times10^{-7}$ $M_\odot$ yr$^{-1}$, the impact is direct because lower mass loss make stars remove less envelope and therefore remain more massive and less chemically enriched at their surface at the end of their main sequence (MS) phase.
For the less massive models ($20$ and $25$ $M_\odot$) with $\dot M\lesssim2\times10^{-8}$ $M_\odot$ yr$^{-1}$, the impact is indirect because lower mass loss make the stars keep high rotational velocities for a longer period of time, then extending the H-core burning lifetime and subsequently reaching the end of the MS with higher surface enrichment.
In either case, given that the conditions at the end of the H-core burning change, the stars will lose more mass during their He-core burning stages anyways.
For the case of $M_\text{zams}=20$ to $40$ $M_\odot$, our models predict stars will evolve through the Hertzsprung gap, from O-type supergiants to blue supergiants and finally red supergiants, with larger mass fractions of helium compared to old evolution models.
New models also sets down to $M_\text{zams}=85\,M_\odot$ the minimal initial mass required for a single star to become a Wolf-Rayet at metallicity $Z=0.002$.}
{These results reinforce the importance of upgrading mass-loss prescriptions in evolution models, in particular for the earlier stages of stellar lifetime, even for $Z\ll Z_\odot$.
New values for $\dot M$ need to be complemented with upgrades in additional features such as convective core overshooting and distribution of rotational velocities, besides more detailed spectroscopical observations from projects such as XShootU, in order to provide a robust framework for the study of massive stars at low metallicity environments.}

\keywords{Stars: evolution -- Stars: massive -- Stars: mass loss -- Stars: rotation -- Stars: winds, outflows}

\maketitle

\section{Introduction}
	The evolution of massive stars (born with $M_\text{zams}\gtrsim10$ $M_\odot$) is important, because they are the progenitors of supernova and of compact objects such as neutron stars or black holes \citep{heger03}.
	Compact objects are of particular interest, because they produce gravitational waves when they are in a binary system and merge.
	At present 90 double compact objects (DCO) mergers have been detected by the LIGO/Virgo/KAGRA collaboration: 76 binary black holes (BH-BH), 12 binary neutron star (NS-NS) and 2 BH-NS mergers \citep{abbott21}.
	Even though the existence of these DCO depends on many considerations regarding the binary system such their common envelope evolution or the binary system spin \citep{olejak21a,olejak21b}, considerations regarding single stellar evolution are also important both for determining the final mass of the compact objects \citep{belczynski10,bavera23,romagnolo24} and for determining the maximal radial expansion of each individual star \citep{agrawal22b,romagnolo23}.
	Moreover, evolution of massive stars is also important for the study of nucleosynthesis, the production of ionising flux, feedback due to wind momentum, for studies of star formation history, and evolution of galaxies.

	Evolution of single massive stars is heavily influenced by their strong stellar winds and subsequently high mass-loss rates ($\dot M$).
	Indeed, changes in the mass loss for the stellar wind considerably affect the evolution of stars with $M_\text{zams}\gtrsim25\,M_\odot$ \citep{meynet94,keszthelyi17}.
	Star models adopting lower values for $\dot M$ retain more mass during their evolution and therefore are larger and more luminous \citet{alex22b,bjorklund23}.
	Besides, weaker winds predict important changes in the evolution of the rotational properties of massive stars \citep{alex23a}, such as a fainter braking in the rotational velocity from the beginning to the end of the main sequence.
	Massive stars keep their high $\varv_\text{rot}$ for a longer fraction of their MS lifetime at the same time they evolve redwards in the HR diagram, before the final braking.
	This appears to be in better agreement with the latest surveys on Galactic O-type stars \citep{holgado22}, even though such survey contains only $285$ stars and thus more observations in the future are required to confirm this trend \citep{alex23a}.

	The adoption of new weaker winds represent an important improvement in our understanding of the evolution of massive stars.	 
	 The release of the X-Shooting Ulysses project \citep[][hereafter XShootU]{vink23a,hawcroft23}, which consist in the observation of several individual massive stars at low metallicity environments, opens a unique opportunity to test new updates in stellar evolution beyond the Milky Way.
	 By spectral modelling, we can determine effective temperature, surface gravity, luminosity, abundances, and mass-loss rate as a function of the metallicity.
	 Even though the majority of the spectral catalog of XShootU is still focused on the Large and Small Magellanic Clouds, they represent an important upgrade towards our understanding of the spectra of the first stellar generations to be provided by the James Webb Stellar Telescope.
	 In order to complement these observations, we additionally need upgraded stellar evolution models.
	 
	 In this paper, we introduce new evolution models for a sample of rotating stars at SMC metallicity ($Z=0.002$).
	 This paper is an extension of our previous works from \citet{alex22b,alex23a}, where we developed stellar evolution models with rotation for Galactic ($Z=0.014$) and LMC ($Z=0.006$), together with evolution models without rotation for the three mentioned metallicities.
	 For that purpose, we use the Geneva-evolution-code \citet[][hereafter \textsc{Genec}]{eggenberger08}, replacing the classical mass-loss prescription from Vink's formula \citep[which has been extensively used for different studies about stellar evolution at different metallicities]{vink01} by our new self-consistent m-CAK prescription \citep{alex19,alex22a} which provides values for $\dot M$ $\sim3$ times lower than the classical winds adopted by \citet{georgy13}.

\section{Physical parameters for evolution models}
\subsection{Self-consistent wind prescription}
	For a solar metallicity environment such as the Milky Way (which is by far the most studied case), we know that the wind structure in massive stars is not homogeneous but clumped, thereby reducing the actual mass-loss rate by a factor of $\sim2-3$ \citep{bouret05,bouret12,surlan12,surlan13}.
	On this basis, important advances have been made in order to both understand in detail the nature of the clumping \citep{sundqvist18,driessen19,driessen22}, and to develop new theoretical prescriptions for the winds of massive stars capable to provide lower values for mass-loss rate, particularly seeking a coherence between the radiative acceleration and the hydrodynamics of the wind.
	Intensive studies such as \citet{sander17} and \citet{alex21} searched for a self-consistent wind solution solving formally the radiative transfer equation in a full NLTE scenario, even though the large computational time required constrained these complex analyses to only a couple of O-type stars.
	Complementary approaches, relaxing some of the particular physical conditions for the atmospheres of massive stars, have succeed into adequately describing the wind hydrodynamics of massive stars, and therefore provide new recipes for the theoretical mass-loss rate for massive stars in a wide range of temperatures and luminosities \citep{kk17,kk18,alex19,alex22a,sundqvist19}, in agreement with empirical spectroscopical analyses \citep{hawcroft21}.
	
	By means of the line-force parameters $k$, $\alpha$, and $\delta$ from the CAK line-driven theory \citep{cak,abbott82,ppk}, \citet{alex19} calculated the line-acceleration simultaneously with solving the equation of motion for the wind hydrodynamics \citep{michel04,michel07}.
	The convergence of this so-called \textit{m-CAK prescription} provides a unique self-consistent mass-loss rate, hereafter $\dot M_\text{sc}$, of the same order of magnitude of the analogous studies of \citet{kk17,kk18} and \citet{sundqvist19}, and in agreement with the observed spectral features for O-type stars \citep{alex22a}, for stars with $T_\text{eff}\ge30$ kK and $\log g\ge3.0$.
	Hence, we implement a recipe for mass-loss rate as a function of a set of stellar parameters, derived from a large grid of line-force parameters for a wide range of stellar masses and metallicities \citep{alex22b}, namely
	\begin{align}\label{mdotalex22b}
		\log\dot M_\text{sc}=&-40.314 + 15.438\,w + 45.838\,x - 8.284\,w\,x \nonumber\\
		&+ 1.0564\,y -  w\,y / 2.36 - 1.1967\,x\,y + 11.6\, z \nonumber\\
		&- 4.223\,w\,z - 16.377\,x\,z + y\,z / 81.735\;,
	\end{align}
	where $\dot M_\text{sc}$ is in $M_\odot$ yr$^{-1}$; and $w$, $x$, $y,$ and $z$ are defined as
	\begin{equation}
		w=\log \left(\frac{T_\text{eff}}{\text{kK}}\right)\;,\nonumber\\
		x=\frac{1}{\log g}\;,\\
		y=\frac{R_*}{R_\odot}\;,\\
		z=\log \left(\frac{Z_*}{Z_\odot}\right)\;.
	\end{equation}
	
	This intervariable fitting $\dot M_\text{sc}(w,x,y,z)$ is achieved by minimising the separation with respect to the resulting mass-loss rate from the $(k,\alpha,\delta)$ calculations, and therefore it depends only on the stellar parameters.
	This is an important difference with respect to the so-called Vink's formula \citep{vink01}, which needs to assume a fixed ratio between the wind terminal velocity $\varv_\infty$ and the escape velocity at the stellar surface $\varv_\text{esc}$
	\footnote{Vink's formula is defined as
	\begin{align}
		\log\dot M_\text{Vink}=&C_1+C_2\log\left(\frac{L_*}{10^5L_\odot}\right)+C_3\log\left(\frac{M_*}{30M_\odot}\right)+C_4\log\left(\frac{\varv_\infty}{2\varv_\text{esc}}\right)\nonumber\\
		&+C_5\log\left(\frac{T_\text{eff}}{20\text{kK}}\right)-C_6\left[\log\left(\frac{T_\text{eff}}{40\text{kK}}\right)\right]^2+0.85\log(Z_*/Z_\odot)\,.\nonumber
	\end{align}
	Where the constants $C_1$, $C_2$, $C_3$, $C_4$, $C_5$, and $C_6$ have different values depending on what temperature range the formula is evaluated.}.
	Likewise, Eq.~\ref{mdotalex22b} can be simplified as
	\begin{align}\label{mdotalex22bnew}
		\log\dot M_\text{sc}=&-40.314 + 15.438\,w + 45.838\,x - 8.284\,w\,x \nonumber\\
		&+ 1.0564\,y -  w\,y / 2.36 - 1.1967\,x\,y\nonumber\\
		&+\log \left(\frac{Z_*}{Z_\odot}\right)\times\left[0.4+\frac{15.75}{M_*/M_\odot}\right]\;,
	\end{align}
	
	For Eq.~\ref{mdotalex22bnew}, metallicity dependent terms have been rewritten as intrinsically dependent on mass, as done in \citet[][see Section~2]{alex23a}, to explicitly emphasise that the mass loss dependence on metallicity does not follow the classical exponential law $\dot M\propto Z^a$ with a constant exponent as normally found in literature \citep{vink21}.
	On the contrary, the influence of metallicity becomes less steep for more massive (and more luminous) stars, which is expected because the contribution of the continuum to the radiative acceleration becomes more important the more luminous is the star \citep{grafener08,kk18,bjorklund21}, and thus the influence of the strongly metallicity-dependent line-acceleration $g_\text{line}=g_\text{rad}-g_\text{cont}$ becomes less relevant.
	
\begin{figure}[t!]
		\centering
		\includegraphics[width=0.9\linewidth]{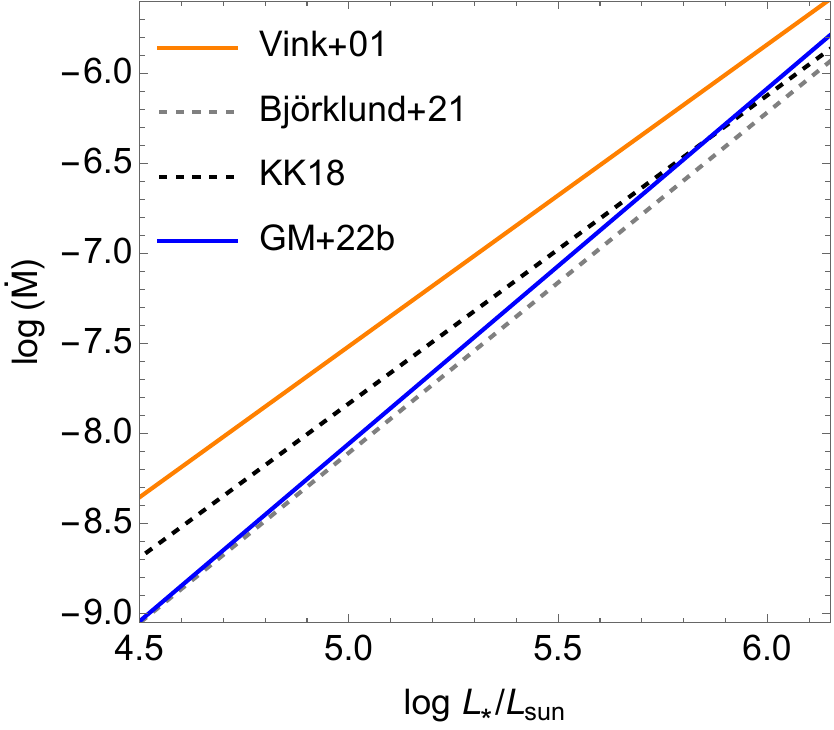}
		\caption{\small{Mass-loss rate as a function of the stellar luminosity at $Z=0.002$, according to different wind prescriptions.}}
		\label{Mdot_vs_lum_SMC}
	\end{figure}

	In Fig.~\ref{Mdot_vs_lum_SMC} we compare our mass-loss recipe \citep{alex22b} with some alternative prescriptions \citep{kk18,bjorklund21} and with \citet{vink01}, all of them for $Z=0.002$.
	New mass-loss prescriptions, including ours, give values $\sim3$ times ($\sim0.5$ in log scale) lower than the standard \citet{vink01} recipe.
	However, $\dot M_\text{sc}$ is closer to Vink's formula at higher luminosities.
	This is a direct consequence of the different metallicity dependences found for both recipes, one being fixed as $\dot M_\text{Vink}\propto Z^{0.85}$ whereas $\dot M_\text{sc}$ becomes shallower at high masses according to Eq.~\ref{mdotalex22bnew}.
	This suggests that m-CAK prescription should provide even stronger winds than Vink's formula for stars massive enough at extremely low metallicities, but that regime is beyond the scope of this work.

\subsection{Other parameters}
	Besides the upgrade of the mass loss recipe for OB-winds during the main sequence, the other physical ingredients for our evolution models are the same as those implemented by \citet{georgy13}.
	This includes, initial rotation $\varv_\text{rot}/\varv_\text{crit}=0.4$, mass loss correction due to rotation following \citet{maeder00}, abundances from \citet{asplund05,asplund09}, opacities from \citet{iglesias96}, and an overshooting parameter of $\alpha_\text{ov}=0.1$.
	Angular momentum transport is described by \citet{zahn92}, with the coefficient $D_\text{eff}$ for the combined horizontal diffusion and meridional currents taken from \citet{chaboyer92} and the $D_\text{shear}$ coefficient for shear turbulence determined by \citet{maeder97}.
	Convective boundaries follow the Schwarzschild criterion.
	We keep these structures to concentrate our analysis on the differences in mass loss over the main sequence only, regardless of studies suggesting larger overshooting values or Tayler-Spruit for the angular momentum transport \citep[see, for instance][]{romagnolo24}.
	
	Mass-loss recipes outside the range of validity of our $\dot M_\text{sc}$ (i.e., for $T_\text{eff}<30$ kK and $\log g<3.0$) are the same as the used by \citet{georgy13}: \citet{vink01} for $T_\text{eff}\ge10$ kK and \citet{dejager88} for $T_\text{eff}<10$ kK.
	For WR winds ($X_\text{surf}\le0.3$ and $T_\text{eff}>10$ kK), we use the formula from \citet{nugis00}.

	\begin{figure*}[t!]
		\centering
		\includegraphics[width=0.6\linewidth]{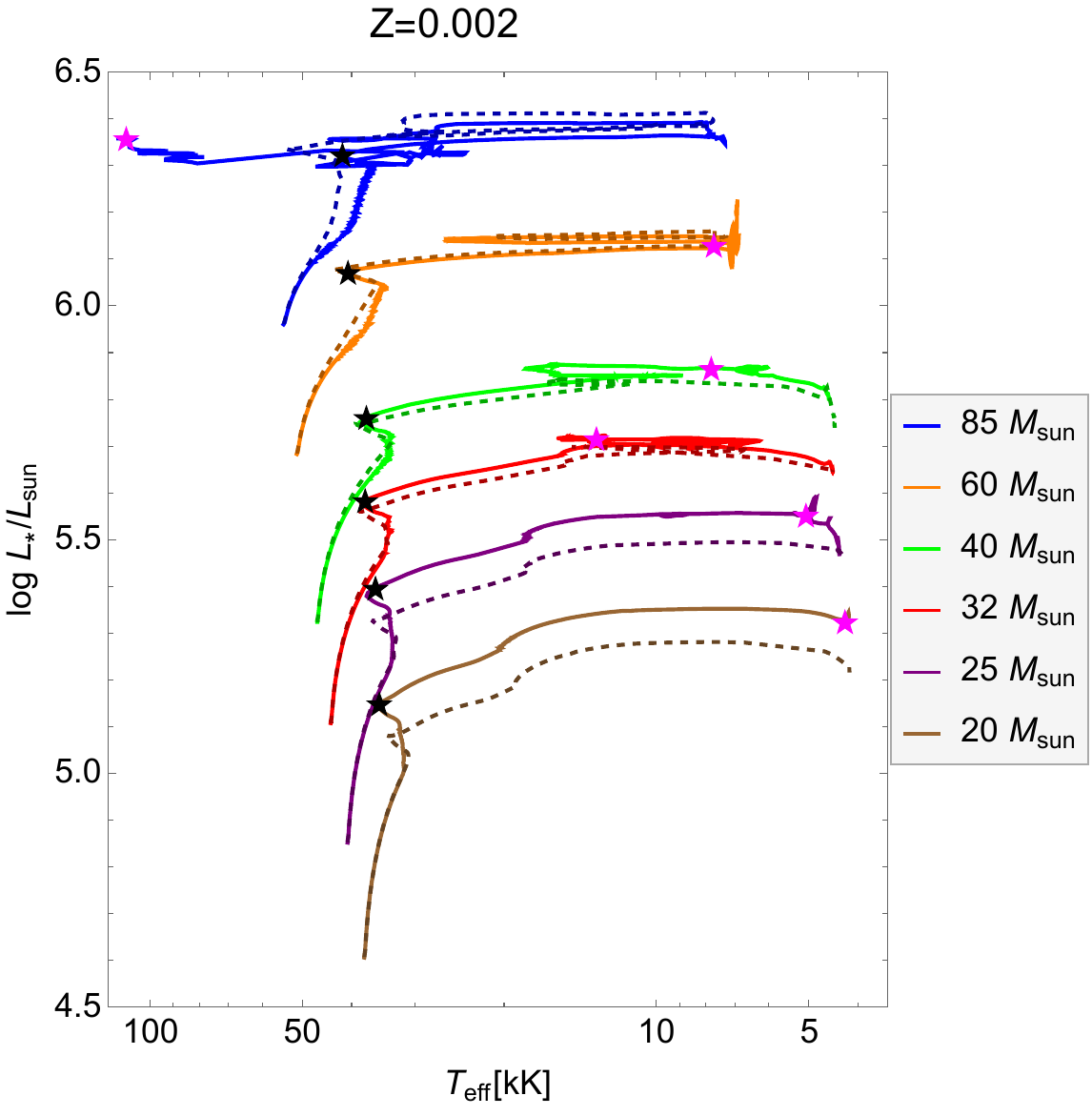}
		\caption{\small{Tracks across HR diagram for evolution models adopting $M_\text{sc}$ (solid lines), compared with tracks adopting $M_\text{Vink}$ from \citet[][dashed lines]{georgy13}.
		New wind prescription gives values for $\dot M$ around $\sim2-6$ times lower than the mass-loss adopted for old models, as visualised in Fig.~\ref{mdots_z02_fin}.
		Black stars mark the end of the H-core burning stage, whereas magenta stars mark the end of the He-core burning stage.}}
		\label{HRD_z02_fin}
	\end{figure*}
\section{Results}
	\begin{table*}[t!]
		\centering
		\caption{Properties of our selected evolutionary tracks at the end of the H-core burning stage, with $\epsilon(\text{N})=\log(\text{N}/\text{H})+12$.
		Values for tracks adopting the Vink's formula are taken from \citet{georgy13}.}
		\begin{tabular}{cc|ccccccccccc}
			\hline
			\hline
			$M_\text{zams}$ & wind recipe & $\varv_\text{rot,ini}$ & $\langle\varv_\text{rot}\rangle$ & $\varv_\text{rot,final}$ & $\tau_\text{MS}$ & $M_\text{total}$ & $M_\text{He-core}$ & $Y_\text{surf}$ & N/C & N/O & $\epsilon$(N) & $R_\text{max}$\\
			$[M_\odot]$ & & \multicolumn{3}{c}{[km s$^{-1}$]} & [Myr] & $[M_\odot]$ & $[M_\odot]$ & \multicolumn{3}{c}{mass fraction} & by number & $[R_\odot]$\\
			\hline
			$85$ & $\dot M_\text{sc}$ & $409.0$ & $323.1$ & $87.2$ & $3.600$ & $80.07$ & $34.92$ & $0.470$ & $5.726$ & $2.334$ & $8.011$ & $35.2$\\
			$85$ & $\dot M_\text{Vink}$ & $435.0$ & $319.0$ & $26.0$ & $3.594$ & $77.24$ & $38.39$ & $0.538$ & $9.794$ & $3.837$ & $8.136$ & $26.1$\\
			\\
			$60$ & $\dot M_\text{sc}$ & $379.0$ & $354.7$ & $111.0$ & $4.192$ & $57.33$ & $20.42$ & $0.378$ & $3.482$ & $1.294$ & $7.837$ & $29.1$\\
			$60$ & $\dot M_\text{Vink}$ & $400.0$ & $317.0$ & $85.0$ & $4.197$ & $56.34$ & $21.77$ & $0.402$ & $4.740$ & $1.648$& $7.901$ & $27.5$\\
			\\
			$40$ & $\dot M_\text{sc}$ & $348.0$ & $323.1$ & $210.0$ & $5.398$ & $39.06$ & $8.43$ & $0.327$ & $3.163$ & $0.950$ & $7.745$ & $21.7$\\
			$40$ & $\dot M_\text{Vink}$ & $358.0$ & $300.0$ & $125.0$ & $5.253$ & $38.67$ & $9.81$ & $0.314$ & $3.218$ & $0.916$ & $7.725$ & $21.4$\\
			\\
			$32$ & $\dot M_\text{sc}$ & $331.0$ & $299.1$ & $258.0$ & $6.624$ & $31.52$ & $4.83$ & $0.336$ & $4.478$ & $1.053$ & $7.791$ & $17.5$\\
			$32$ & $\dot M_\text{Vink}$ & $338.0$ & $282.0$ & $177.0$ & $6.354$ & $31.24$ & $5.74$ & $0.310$ & $4.072$ & $0.927$ & $7.739$ & $16.4$\\
			\\
			$25$ & $\dot M_\text{sc}$ & $314.0$ & $272.9$ & $285.0$ & $8.551$ & $24.79$ & $1.16$ & $0.356$ & $7.380$ & $1.205$ & $7.849$ & $13.9$\\
			$25$ & $\dot M_\text{Vink}$ & $319.0$ & $262.0$ & $132.0$ & $7.663$ & $24.67$ & $2.04$ & $0.292$ & $4.655$ & $0.862$ & $7.721$ & $13.8$\\
			\\
			$20$ & $\dot M_\text{sc}$ & $300.0$ & $252.0$ & $277.0$ & $10.513$ & $19.93$ & $0.12$ & $0.336$ & $8.849$ & $1.143$ & $7.833$ & $11.5$\\
			$20$ & $\dot M_\text{Vink}$ & $305.0$ & $248.0$ & $194.0$ & $9.238$ & $19.87$ & $0.11$ & $0.277$ & $4.736$ & $0.792$ & $7.697$ & $11.6$\\
			\hline
		\end{tabular}
		\label{table_MS_z02}
	\end{table*}
	\begin{figure*}[t!]
		\centering
		\hspace{0.2mm}
		\includegraphics[width=0.7\linewidth]{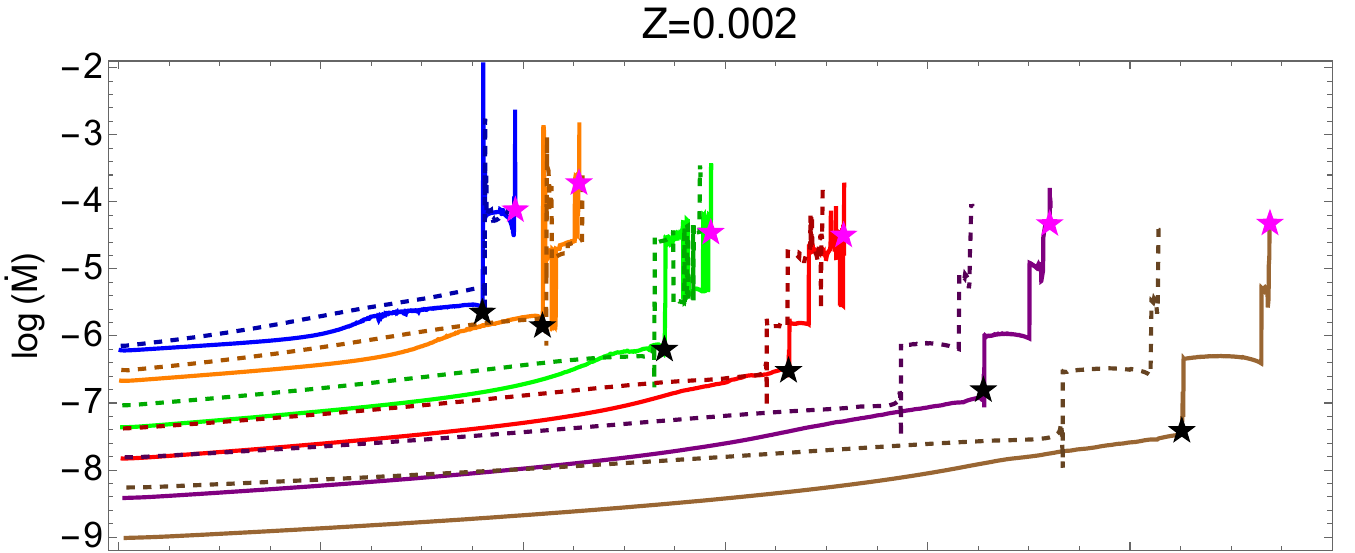}\\
		\vspace{-0.5mm}
		\includegraphics[width=0.706\linewidth]{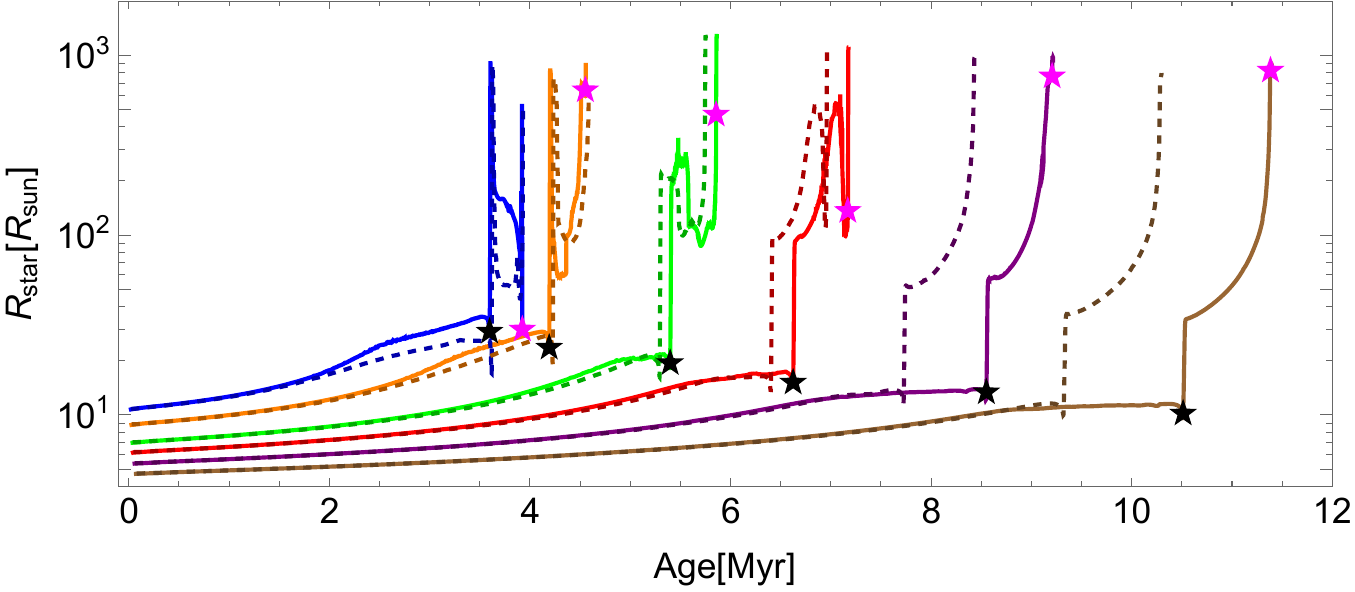}		
		\caption{\small{Mass-loss rates (upper panel) and stellar radius (lower panel) as a function of time, for our new $\dot M_\text{sc}$ prescription (solid lines) and for $\dot M_\text{Vink}$ (dashed lines).
		The legends for the colours and the black and magenta stars are the same as for Fig.~\ref{HRD_z02_fin} }}
		\label{mdots_z02_fin}
	\end{figure*}
	\begin{figure*}[t!]
		\centering
		\includegraphics[width=0.4\linewidth]{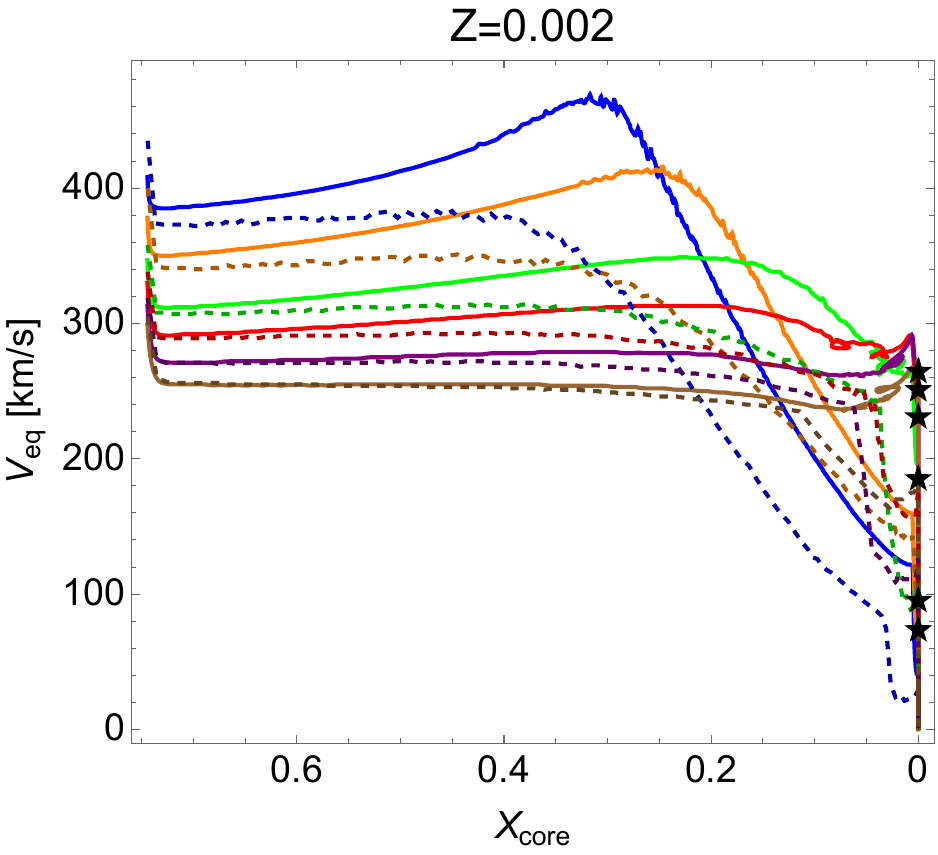}
		\hspace{1cm}
		\includegraphics[width=0.4\linewidth]{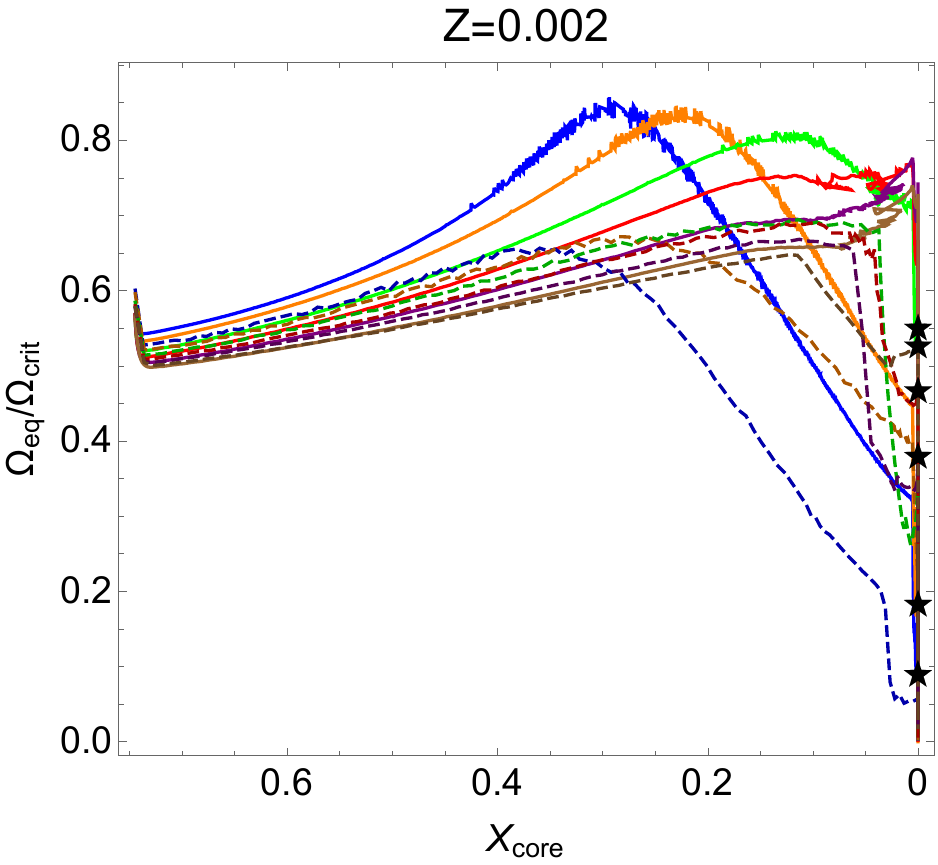}
		\caption{\small{Evolution of the rotational velocity expressed as the absolute magnitude $\varv_\text{rot}$ (left panel) and the angular velocity as a fraction of the critical velocity $\Omega/\Omega_\text{crit}$ (right panel).
		New models exhibit larger velocities because they lose less angular momentum due to weaker winds.
		The legends for the colours and the black stars are the same as for Fig.~\ref{HRD_z02_fin} and Fig.~\ref{mdots_z02_fin}.}}
		\label{rotation_z02_fin}
	\end{figure*}

	\begin{figure}[t!]
		\centering
		\includegraphics[width=0.9\linewidth]{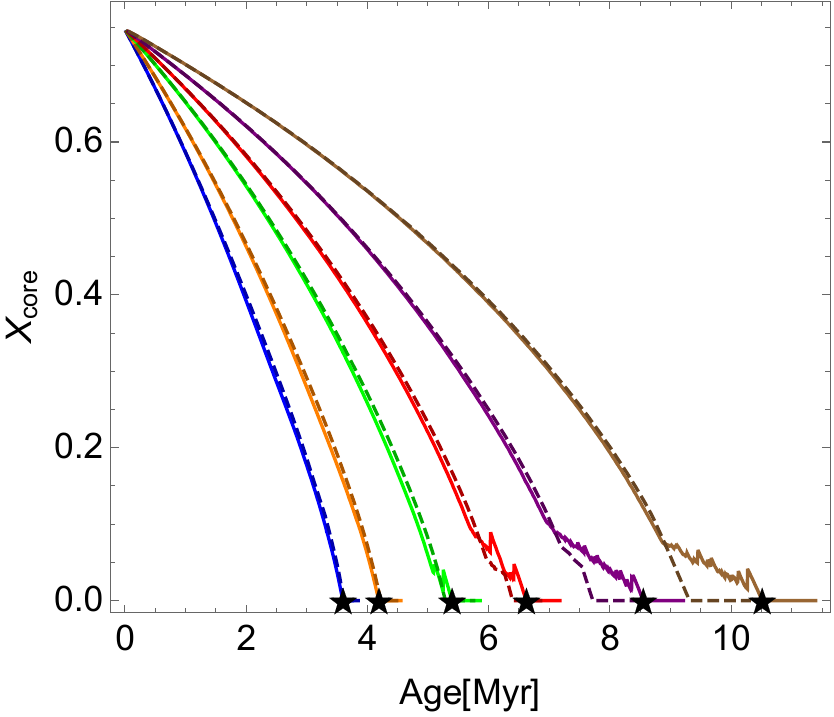}
		\caption{\small{Evolution of the hydrogen mass fraction at the core, showing the extension of the MS lifetime for our less massive models.
		The legends for the colours and the black stars are the same as for previous figures.}}
		\label{Xcore_z02_fin}
	\end{figure}
	\begin{figure*}[t!]
		\centering
		\includegraphics[width=0.33\linewidth]{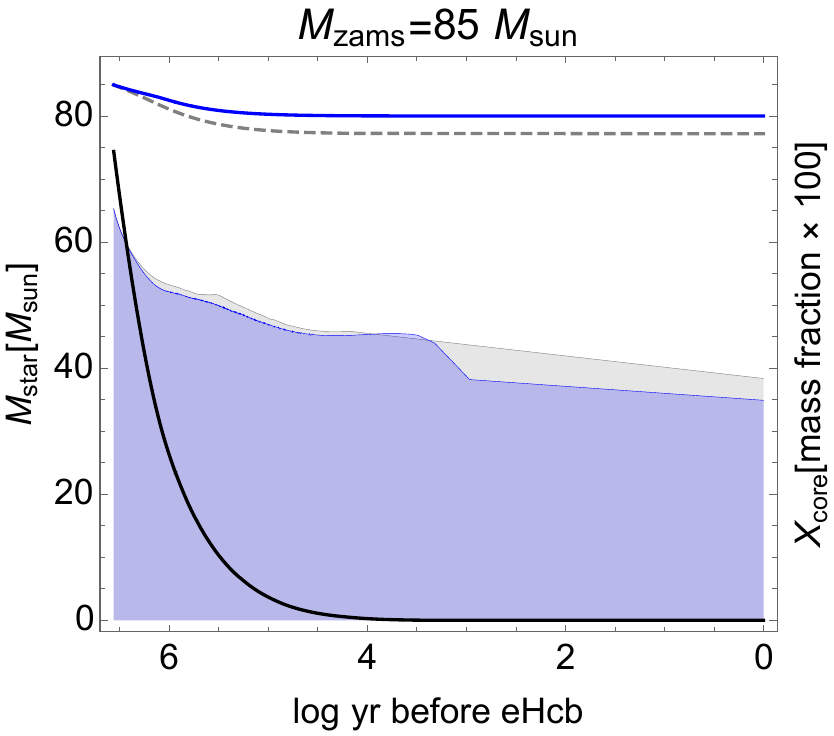}
		\includegraphics[width=0.33\linewidth]{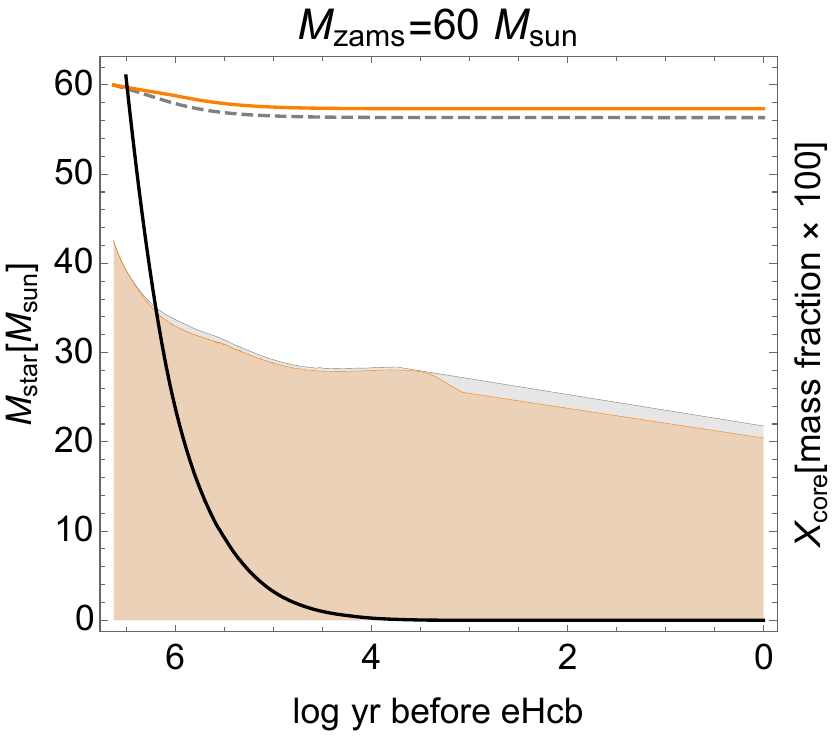}
		\includegraphics[width=0.33\linewidth]{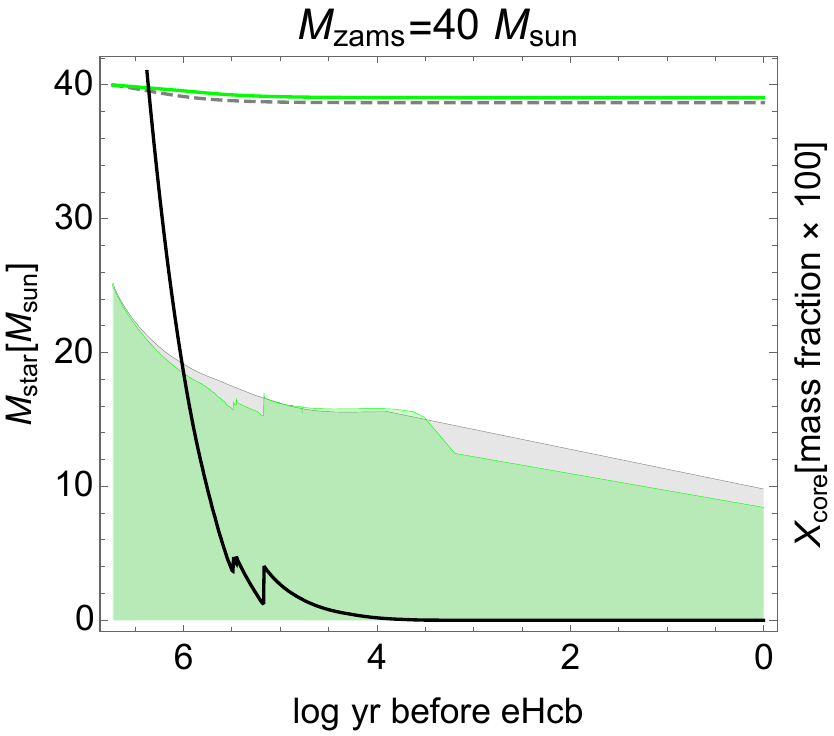}\\
		\includegraphics[width=0.33\linewidth]{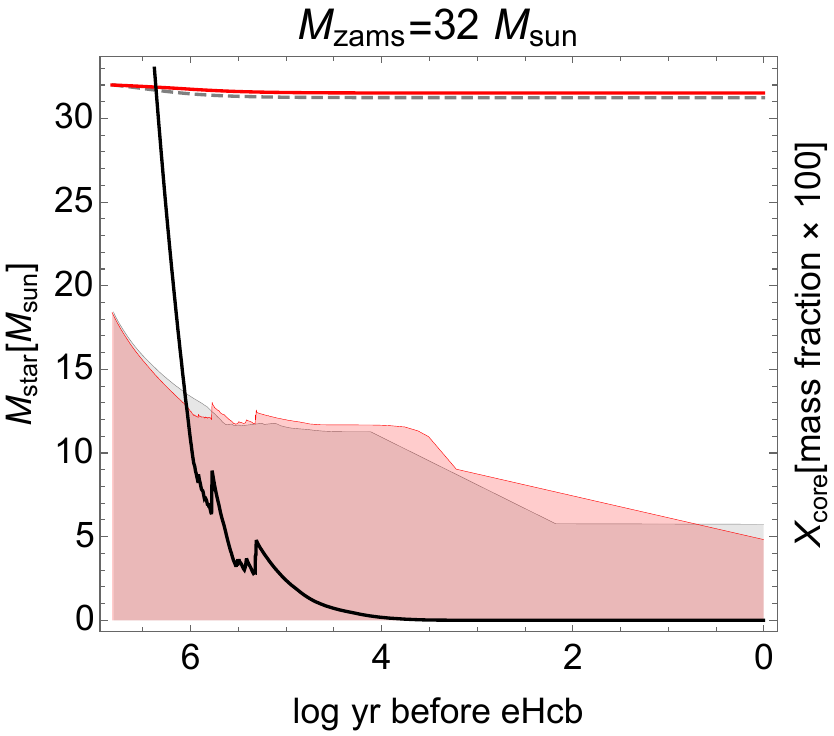}
		\includegraphics[width=0.33\linewidth]{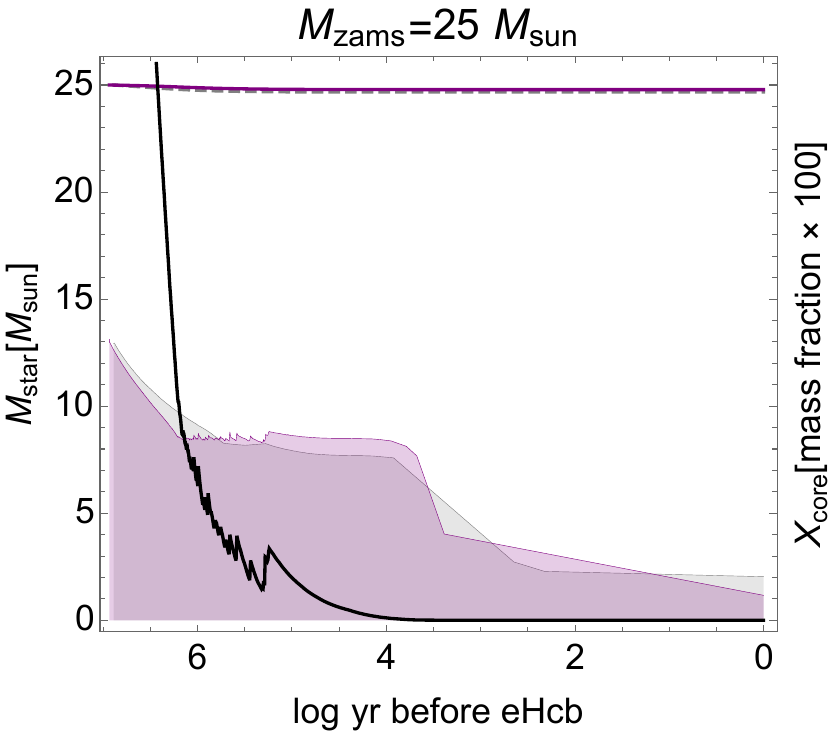}
		\includegraphics[width=0.33\linewidth]{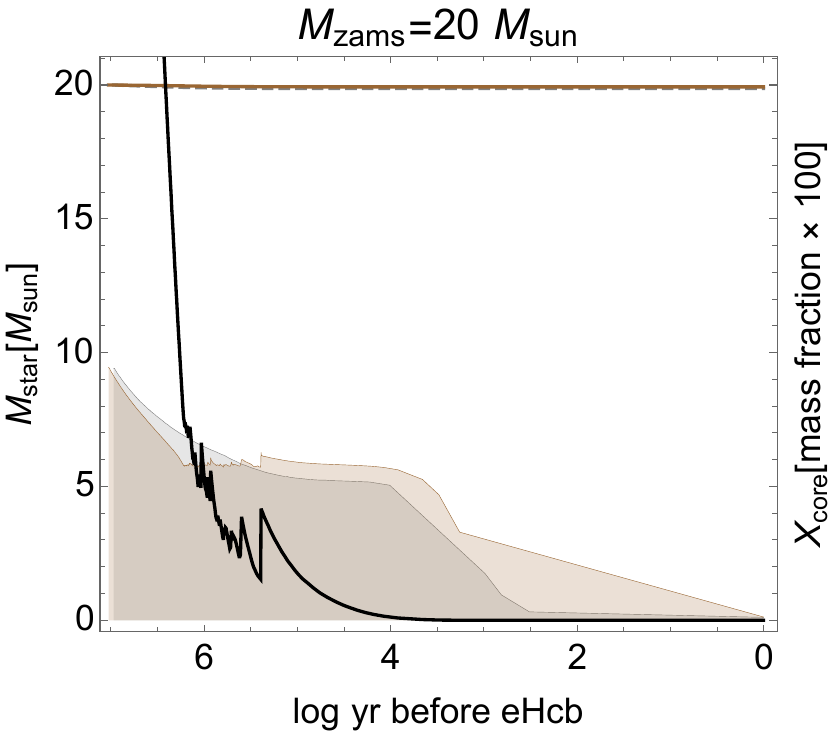}
		\caption{\small{Kippenhahn diagrams, showing the evolution of the stellar total mass and the core mass, for our studied range of masses.
		Coloured solid coloured lines and shadowed areas represent models adopting new self-consistent winds, whereas grey dashed lines and shadowed areas represent models adopting old winds.
		Black tracks correspond to the percentage of hydrogen mass fraction in the core.}}
		\label{kipperhahn_z02}
	\end{figure*}
	
	The tracks for our new evolution models adopting self-consistent mass-loss rates at $Z=0.002$, compared with models adopting old winds from \citet{georgy13}, are shown in Fig.~\ref{HRD_z02_fin}.
	The evolution of the mass-loss rates and the stellar radius, comparing $\dot M_\text{sc}$ and $\dot M_\text{Vink}$, are shown in Fig.~\ref{mdots_z02_fin}.
	The evolution of the rotational velocities is shown in Fig.~\ref{rotation_z02_fin}.
	Table~\ref{table_MS_z02} summarises the stellar properties at the end of the main sequence (H-core burning) stage, whereas Table~\ref{table_HeC_z02} summarises the properties at the end of He-core and C-core burning stages.

\subsection{Evolution through the main sequence}\label{resultsHburn}
	Figure~\ref{mdots_z02_fin} shows that the differences between the mass-loss recipes are narrower for the most massive and luminous stellar models.
	According to Eq.~\ref{mdotalex22bnew}, at $Z=0.002$ ($=0.14Z_\odot$) the difference between $\dot M_\text{sc}$ and $\dot M_\text{Vink}$ is only $\sim0.05$ for our $85\,M_\odot$ model, whereas for our $20\,M_\odot$ model the starting $\dot M_\text{sc}$ is $\sim6.25$ times lower than $\dot M_\text{Vink}$ ($\sim0.8$ dex), as also seen for the non-rotating models at $Z=0.002$ \citep{alex22b}.
	However, even though the comparative difference in $\dot M$ is smaller for our most massive models ($60$ and $85$ $M_\odot$), the upgrade in the mass-loss rate still has an important impact over their evolutionary tracks because their values are the highest in absolute terms ($\gtrsim2\times10^{-7}$ $M_\odot$ yr$^{-1}$).
	Such impact is the same observed in \citet{alex23a} but on a minor scale: weaker wind makes the star rotate faster (Fig.~\ref{rotation_z02_fin}) whereas it keeps more of its envelope, which results in a drift redwards in the HR diagram because the inner structure is less homogeneous.
	Homogeneous chemical composition in the inner stellar structure favours bluewards evolution, as demonstrated by \citet{maeder87c}.
	Consequently, we find for new evolution models a larger rotational speed during all their MS lifetime but particularly at the end of it, and a more moderate change in the surface abundances in comparison with old tracks (Table~\ref{table_MS_z02}).
	
	On the other side, for the less massive models, comparative differences in the $\dot M$ recipe are larger but the absolute magnitude of it is smaller ($\lesssim2\times10^{-8}$ $M_\odot$ yr$^{-1}$).
	So this time the impact over the stellar evolution is minimal, at least during most of the main sequence.
	Still, in this case the weaker wind makes the star stay in the MS a nearly $12\%$ longer time for our $25\,M_\odot$ model, and about $14\%$ longer time for our $20\,M_\odot$ model.
	As a consequence, changes in the final surface abundances at the end of H-core burning are larger for the new evolution models adopting weaker winds, because envelope removal becomes negligible and then the changes in the surface composition are preponderantly influenced by the rotational mixing \citep{maeder00,maeder10}.
	
	The MS lifetime is enhanced for our models with $M_\text{zams}=20$, $25$, $32$, and (though on a minor scale) $40$ $M_\odot$.
	These are also the models ending their H-core burning stages with the highest rotational velocities, over $200$ km s$^{-1}$ and angular momentum $\Omega/\Omega_\text{crit}\gtrsim0.7$ according to Table~\ref{table_MS_z02} and Fig.~\ref{rotation_z02_fin}.
	Although in our case we are just analysing $Z=0.002$, this trend agrees with \citet{szecsi15} and \citet{kubatova19}, where stars at low metallicities keep faster rotations due to their weaker winds.
	Because of this rapid rotation, the inner rotational mixing is strong enough to add more hydrogen to the convective core, thus making the stellar core to have more fuel to burn and therefore extend the H-core burning lifetime.
	Indeed, in Fig.~\ref{Xcore_z02_fin} we can see how the hydrogen abundance in the core is `bumped' prior to its total depletion for our new evolution models.
	The same behaviour is illustrated with more details in Fig.~\ref{kipperhahn_z02}, which shows Kippenhahn diagrams for our six stars covering their MS lifetimes.
	These bumps in the core hydrogen abundance coincide with the increase in the size of the convective core, meaning that the core gains fuel to extend the H-burning process.
	The exceptions for this phenomenon are the most massive cases, $60$ and $85$ $M_\odot$, where the mass loss is strong enough to brake the star prior to the end of H-core burning, and therefore the inner mixing does not refill the convective core.
	
\begin{figure*}[t!]
		\centering
		\includegraphics[width=0.45\linewidth]{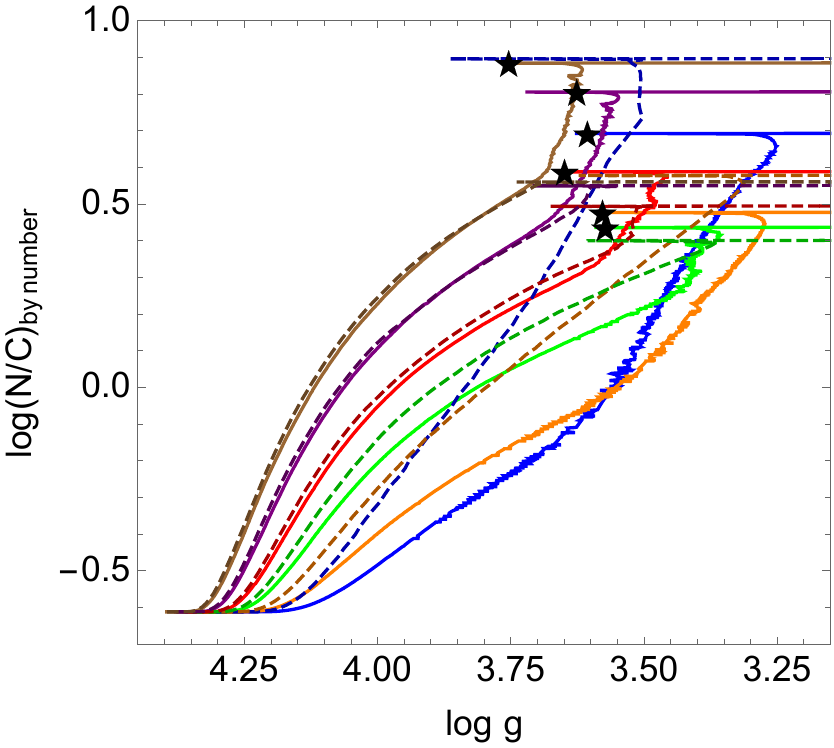}
		\hspace{1cm}
		\includegraphics[width=0.45\linewidth]{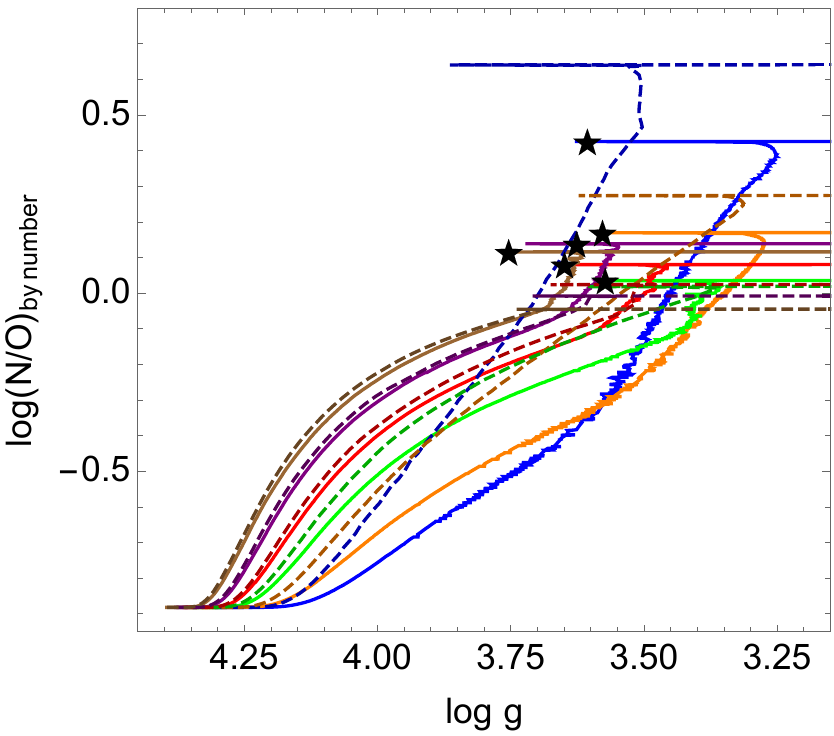}
		\caption{\small{Evolution of the ratios of the abundance fractions N/C (left panel) and N/O (right panel).
		The legends for the colours and the black crosses are the same as for previous figures.}}
		\label{NCNOlogg_z02_fin}
	\end{figure*}

	The new recipe of mass-loss rate has a direct and indirect impacts on the stellar evolution, depending on the initial stellar mass.
	The impact over the stellar evolution at SMC is direct for the most massive stars ($60$ and $85$ $M_\odot$), because weaker winds remove less material from the outer stellar envelope, thus favouring a less homogeneous structure and subsequently a more redwards evolution.
	On the other side, for the less massive stars of our grid the impact is indirect, and it is observed by means of a more efficient mixing due to a more rapid rotation and a more extended MS lifetime.
	The contrast between both direct and indirect impacts, which is actually the struggle between mass loss and rotational mixing as the dominant factor driving the evolution of massive stars, is particularly evident in the resulting chemical enrichment at the stellar surface.
	Differences in the resulting CNO abundances at the stellar surfaces are shown in Table~\ref{table_MS_z02} and the N/C and N/O ratios are plotted in Fig.~\ref{NCNOlogg_z02_fin}, where it is straightforward to observe the contrast in the effects for the most and the less massive models.
	The changes in the N/C are more drastic than the N/O ratio due to the conditions of the CNO-cycle \citep{maeder14}.
	
	More massive models do not substantially change their MS lifetimes, but the final mixing at the surface is reduced (compared with older evolution models) because weaker winds remove less envelope.
	Also, the evolution redwards observed for $60$ and $85$ $M_\odot$ models implies larger radii and thus lower surface gravity.
	As a consequence, for our new models the $\log g$ goes down to $\sim3.3$ before the final contraction prior to the end of the H-core burning stage, and the final N/C and N/O ratios are lower.
	On the contrary, for the less massive stars of our model grid ($20$ and $25$ $M_\odot$), the N/C and N/O barely change with the decreasing of the surface gravity, but because less massive models have more extended MS lifetimes and consequently larger final surface mixings, the abundance of nitrogen continues growing until exceeding the final enrichment shown by the old models.
	In the middle we have the $40\,M_\odot$ and the $32\,M_\odot$ models, as moderate versions of both previous effects.
	The combination of these effects makes our evolutionary tracks to cover wider areas on the planes N/C and N/O-vs-$\log g$ and (Fig.~\ref{NCNOlogg_z02_fin}), showing these plots the biggest differences between old and new winds.
	
	In summary, new evolution models adopting our self-consistent recipe enhance the influence of the rotational mixing on the surface chemical enrichment, in detriment of the influence of the mass loss, which is a logical consequence of their lower $\dot M$ values.
	In particular for $Z_\text{SMC}=0.002$ the stellar mass showing the lowest N/C ratio is around $40$ $M_\odot$, for both old and new winds.
	Over and below that mass, either mass loss or rotational mixing respectively increases nitrogen enrichment \citep{bouret21}.
	However, for the case of N/O we observe that the lowest enrichment ratio belongs to the model with $20\,M_\odot$ adopting old mass-loss recipe \citep{georgy13} and to the model with $40\,M_\odot$ adopting new recipe.
		
	\begin{table*}[t!]
		\centering
		\caption{Properties of our selected evolutionary tracks at the end of the He-core and C-core burning stage.
		Values for tracks adopting the Vink's formula are taken from \citet{georgy13}.}
		\resizebox{\linewidth}{!}{
		\begin{tabular}{cc|cccccc|ccccccc}
			\hline
			\hline
			& & \multicolumn{6}{c}{End of He-core burning} & \multicolumn{6}{c}{End of C-core burning}\\
			$M_\text{zams}$ & wind recipe & $\tau_\text{He}$ & $M_\text{total}$ & $Y_\text{surf,fin}$ & N/C & N/O & $R_\text{max}$ & $\tau_\text{C}$ & $M_\text{total}$ & $Y_\text{surf,fin}$ & N/C & N/O & $R_\text{max}$\\
			$[M_\odot]$ & & [Myr] & $[M_\odot]$ & \multicolumn{3}{c}{mass fraction} & $[R_\odot]$ & [kyr] & $[M_\odot]$ & \multicolumn{3}{c}{mass fraction} & $[R_\odot]$\\
			\hline
			$85$ & $\dot M_\text{sc}$ & 0.325 & 47.4 & 0.599 & 0.339 & 0.128 & 831.0\\
			$85$ & $\dot M_\text{Vink}$ & 0.305 & 51.36 & 0.826 & 132.2 & 89.00 & 882.4 & 0.002 & 51.13 & 0.827 & 128.9 & 87.72 & 536.6\\
			\\
			$60$ & $\dot M_\text{sc}$ & 0.361 & 33.14 & 0.809 & 65.45 & 91.32 & 870.4 & 0.053 & 31.82 & 0.809 & 93.97 & 89.24 & 910.7\\
			$60$ & $\dot M_\text{Vink}$ & 0.350 & 39.45 & 0.744 & 64.00 & 33.75 & 744.8 & 0.006 & 39.12 & 0.754 & 70.00 & 50.91 & 682.9\\
			\\
			$40$ & $\dot M_\text{sc}$ & 0.458 & 30.24 & 0.626 & 27.02 & 8.35 & 466.2 & 0.061 & 28.74 & 0.662 & 38.35 & 16.44 & 1280.7\\
			$40$ & $\dot M_\text{Vink}$ & 0.448 & 29.16 & 0.572 & 20.00 & 4.984 & 966.6 & 0.045 & 27.44 & 0.637 & 39.42 & 12.96 & 1288.3\\
			\\
			$32$ & $\dot M_\text{sc}$ & 0.545 & 24.53 & 0.614 & 31.58 & 7.711 & 611.6 & 0.127 & 23.95 & 0.636 & 39.46 & 12.02 & 1106.1\\
			$32$ & $\dot M_\text{Vink}$ & 0.543 & 24.68 & 0.578 & 28.16 & 5.494 & 518.8 & 0.107 & 24.07 & 0.597 & 33.56 & 7.061 & 1103.6\\
			\\
			$25$ & $\dot M_\text{sc}$ & 0.655 & 20.04 & 0.435 & 11.97 & 1.677 & 898.7 & 0.213 & 19.11 & 0.616 & 43.62 & 10.66 & 984.2\\
			$25$ & $\dot M_\text{Vink}$ & 0.694 & 22.77 & 0.316 & 6.372 & 1.036 & 964.3 & 0.256 & 21.83 & 0.322 & 6.812 & 1.077 & 987.3\\
			\\
			$20$ & $\dot M_\text{sc}$ & 0.871 & 18.88 & 0.358 & 10.55 & 1.25 & 834.0 & 0.576 & 18.19 & 0.371 & 11.75 & 1.32 & 899.4\\
			$20$ & $\dot M_\text{Vink}$ & 0.945 & 19.25 & 0.285 & 5.663 & 0.865 & 710.0 & 0.630 & 18.66 & 0.302 & 7.589 & 0.999 & 793.0\\
			\hline
		\end{tabular}}
		\label{table_HeC_z02}
	\end{table*}
	
\subsection{Evolution through later stages}
	Beyond the main sequence, for He-core and C-core burning stages, mass loss is described by the same recipes implemented by the models of \citet{georgy13}: Vink's formula if $X_\text{s}\ge0.3$, or \citet{grafener08} if $X_\text{sc}\le0.3$.
	However, because the changes in the MS mass loss recipes make evolution models finish their H-core burning stages at different locations of the HRD and with different rotational and chemical properties, their subsequent evolution will retain important differences.
	
	Table~\ref{table_HeC_z02} summarises the final properties of our evolution models at the end of He-core and C-core burning, in comparison with the models of \citet{georgy13}.
	Here again, the contrast between the models dominated by mass loss and the models dominated by rotational mixing becomes even stronger.
	For our masses $20$, $25$ and $32$ $M_\odot$, which were the models whose MS lifetime was extended and more chemically evolved at the end of H-core burning, the final masses are smaller than their counterpart models adopting old winds.
	This is due to the fact that these models expand redwards with larger luminosities and larger rotation, thus with larger mass-loss rates, also implying that the CNO mixing at the surface is more enhanced.
	These effects are less relevant for our $40$ $M_\odot$ model, where the final abundances at the end of the MS changed the less when we adopt the new winds.
	In all of these four cases, the stars continue their expansion post-MS even after the end of their He-core burning stages, reason why their largest radii are found during the C-core burning stage when these stars become RSG.
	The RSG predicted by our $25$ to $40$ $M_\odot$ models satisfy $\log L_*/L_\odot\lesssim5.7$, thus being in agreement with the empirical results from \citet{davies18}, where no RSG have been found for higher luminosities.
	The maximum radial expansion of these RSG is barely affected (only our $20$ $M_\odot$ increases its maximum radius by $\sim100\,R_\odot$), thus making $R_\text{max}$ grow with mass from $\sim900\,R_\odot$ at $M_\text{zams}=20\,M_\odot$ to $\sim1280\,R_\odot$ at $M_\text{zams}=40\,M_\odot$.
	That maximum radial expansion (for all our masses) lies close of the $\sim1300\,R_\odot$ found for the RSG Dachs SMC 1–4 \citep[the largest star known in the SMC,][]{gilkis21}, but its determined luminosity $\log L_\text{Dachs SMC 1–4}/L_\odot\simeq5.55$ does not coincide with the luminosity predicted by our $40\,M_\odot$ model ($L_*/L_\odot\simeq5.7$).
	This reinforces that, besides the upgrade on mass loss, evolution models need to revise the convection treatment for the inner layers, given that a less efficient semiconvective mixing extends the band on the HRD where RSG can exist \citep{schootemeijer19}.
	Luminosity of red supergiants is also important, because it is directly correlated with the $M_\text{He,core}$ prior to the core-collapse supernovae \citep{farrell20}.
	
	On the other side, our $60$ and $85$ $M_\odot$ models end the H-core burning stage with more mass and a less chemically evolved surface (compared with models for the same masses adopting old winds) prior to the post-MS expansion.
	Because they also reach with larger masses at the end of that radial expansion (i.e., when they reach $T_\text{eff}\le9$ kK and then return bluewards), they experience a mass loss large enough to revert the previous scenario and then end their He-core burning stages with smaller masses.
	Let us remind that, for these stars, $\dot M\gtrsim10^{-6}$ $M_\odot$ yr$^{-1}$, and therefore the discrepancies in mass loss create a stronger impact.
	For the case of our $60\,M_\odot$ model, this extra mass loss during the He-core burning is expressed not only in a smaller final mass but also in larger N/C and N/O ratios as consequence of the extra removal of envelope.
	The case of our $85\,M_\odot$ model is even more extreme, because this extra removal includes the latest envelopes of hydrogen and therefore the star becomes a Wolf-Rayet, thus following a completely different track in comparison with the corresponding model adopting old wind recipe, namely, larger temperatures and a CNO mass fraction of $\sim40\%$, with reduced N/C and N/O ratios (Table~\ref{table_HeC_z02}) because surface nitrogen has been quickly removed.
	In other words, the adoption of self-consistent winds for $\dot M$ reduces the threshold on $M_\text{zams}$ over the which stars are expected to become WR at $Z=0.002$.
	This is important, because it implies that new evolution models might have a better match with the WR stars at SMC studied by \citet{hainich15}, which could not be reproduced by the evolution models of \citet{georgy13}.
	Nevertheless, we need to the upgrade not only the $\dot M$ recipe for OB-type stars optically thin winds but also for the WR optically thick winds \citep{bestenlehner20,sander20,sander23}, besides revisiting the criteria for the transition between these wind regimes \citep[which depends more on the proximity to the Eddington limit than on the surface hydrogen abundance, as asserted by][]{grafener21}, before a more robust prediction for WR stars can be performed.
	
	We reiterate that the mass-loss recipes adopted during He-core burning stage are the same as in \citet{georgy13}, because self-consistent m-CAK prescription works only for $T_\text{eff}\ge30$ kK.
	Henceforth, the described differences between old and new evolution models stem from the models ending the H-core burning stage with different masses and chemical abundances at the surface.
	This reinforces the importance of upgrading mass-loss formulae for massive stars during their main sequence, specially because our self-consistent $\dot M$ is $\sim2-6$ times lower than the formula from \citet{vink01} originally adopted for the SMC evolution models.

\section{Observational diagnostics}
	\begin{figure*}[t!]
		\centering
		\includegraphics[width=0.47\linewidth]{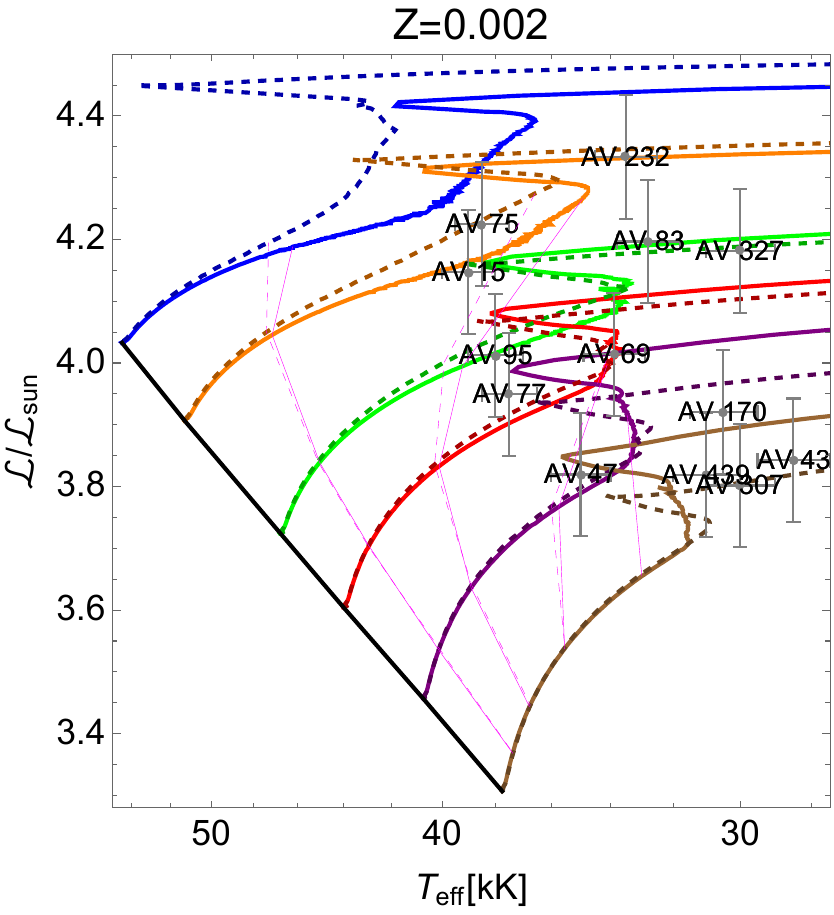}
		\hspace{3mm}
		\includegraphics[width=0.48\linewidth]{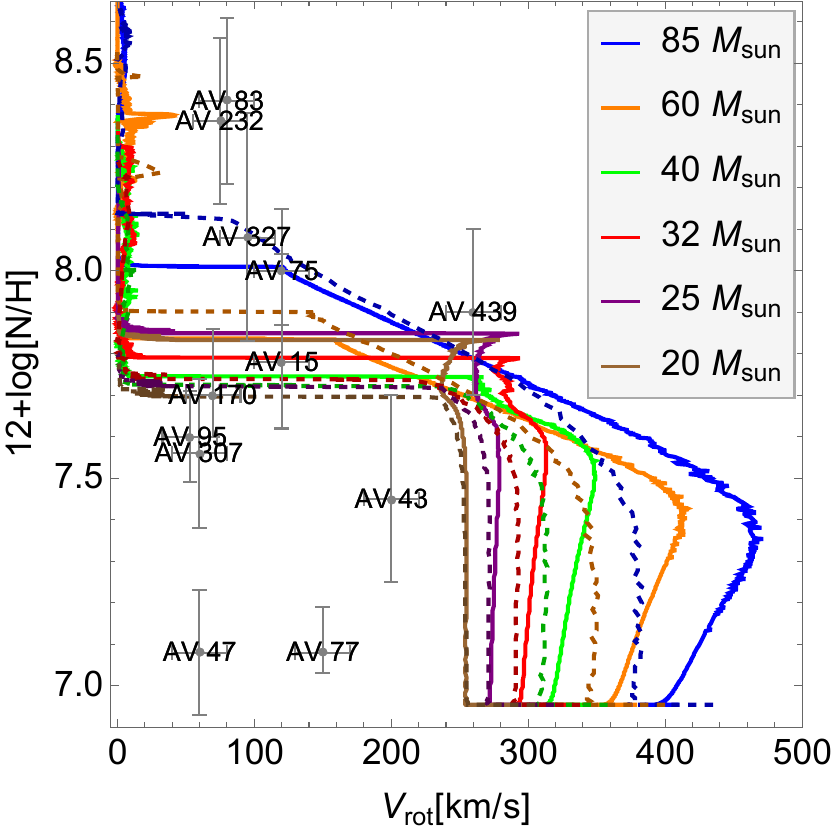}
		\caption{\small{Left panel: spectroscopic HR diagram showing $T_\text{eff}$ vs $\mathcal L(=T_\text{eff}^4/g)$ for new and old evolution models, comparing with SMC evolved stars from \citet[][]{bouret21}: AV 15, AV 43, AV 47, AV 69, AV 75, AV 77, AV 83, AV 95, AV 170, AV 232, AV 307, AV 327, and AV 439.
		The 2, 4, 6, and 8 Myr isochrones for old and new tracks are shown in magenta (dashed and solid lines respectively).
		Right panel: Hunter diagram \citep{hunter09} for new and old evolution models, comparing with the SMC evolved stars from \citet{bouret21}.
		The studied stars are labeled with their names, and their full properties are listed in Table~\ref{stars_bouret}.
		Because of its low N/H, AV 69 lies out of the plot range in the right panel.}}
		\label{SHRD_z02_fin}
	\end{figure*}
	\begin{table*}
		\centering
		\caption{Stars from \citet{bouret21}.
		The helium abundances, originally given by number, were readapted as mass fractions for purposes of this work.}
		\begin{tabular}{ccccccccc}
			\hline
			\hline
			Star & Sp. type & $T_\text{eff}$ & $\log g$ & $L_*$ & $\varv\sin i$ & $Y_\text{He}$ & $\epsilon(\text{N})$ & $\log$(N/C)\\
			& & [kK] & & [$L_\odot$] & [km s$^{-1}$] & mass fraction & \multicolumn{2}{c}{by number}\\
			\hline
			\vspace{1mm}
			AV 15 &  O6.5I(f) & 39.0 & 3.61 & 5.83 & 120.00 & 0.285 & $7.78\pm0.16$ & $0.78_{-0.17}^{+0.23}$\\
			\vspace{1mm}
			AV 43 & B0.5III & 28.5 & 3.37 & 5.13 & 200.00 & 0.374 & $7.45_{-0.20}^{+0.25}$ & $-0.03_{-0.21}^{+0.43}$\\
			\vspace{1mm}
			AV 47 & O8III((f)) & 35.0 & 3.75 & 5.44 & 60.00 & 0.285 & $7.08\pm0.15$ & $-0.61_{-0.15}^{+0.20}$\\
			\vspace{1mm}
			AV 69 & OC7.5III((f) & 33.9 & 3.50 & 5.61 & 70.00 & 0.285 & $6.34\pm0.17$ & $-1.22_{-0.19}^{+0.27}$\\
			\vspace{1mm}
			AV 75 &  O5.5I(f) &  38.5 & 3.51 & 5.94 & 120.00 & 0.324 & $8.00_{-0.13}^{+0.15}$ & $1.00_{-0.15}^{+0.29}$\\
			\vspace{1mm}
			AV 77 & O7III & 37.5 & 3.74 & 5.40 & 150.00 & 0.374 & $7.08_{-0.05}^{+0.11}$ & $-0.10_{-0.15}^{+0.19}$\\
			\vspace{1mm}
			AV 83 & O7Iaf+ & 32.8 & 3.26 & 5.54 & 80.00 & 0.444 & $8.41\pm0.20$ & $0.83_{-0.18}^{+0.28}$\\
			\vspace{1mm}
			AV 95 & O7III((f)) & 38.0 & 3.70 & 5.46 & 53.00 & 0.285 & $7.60\pm0.11$ & $0.30_{-0.13}^{+0.16}$\\
			\vspace{1mm}
			AV 170 & O9.7III & 30.5 & 3.41 & 5.14 & 70.00 & 0.324 & $7.70_{-0.10}^{+0.16}$ & $0.40_{-0.14}^{+0.22}$\\
			\vspace{1mm}
			AV 232 & O7Iaf+ & 33.5 & 3.16 & 5.89 & 75.00 & 0.444 & $8.36\pm0.20$ & $0.86_{-0.18}^{+0.28}$\\
			\vspace{1mm}
			AV 307 & O9III & 30.0 & 3.50 & 5.15 & 60.00 & 0.324 & $7.56\pm0.18$ & $0.18_{-0.20}^{+0.30}$\\
			\vspace{1mm}
			AV 327 & O9.7I & 30.0 & 3.12 & 5.54 & 95.00 & 0.374 & $8.08_{-0.25}^{+0.20}$ & $0.78_{-0.21}^{+0.28}$\\
			\vspace{1mm}
			AV 439 & O9.5III & 31.0 & 3.54 & 5.16 & 260.00 & 0.308 & $7.90\pm0.20$ & $0.28_{-0.20}^{+0.29}$\\
			\hline
		\end{tabular}
		\label{stars_bouret}
	\end{table*}
	
	We compare the predicted stellar evolution for massive stars at SMC metallicity with the observational diagnostics performed \citet{bouret21} for O-type giants and supergiants.
	We do not include O-dwarves from \citet{bouret13}, because they are in a very early stage of the main sequence, where the differences between old and new evolution models are barely perceptible.
	Despite the most remarkable differences between tracks adopting either old or new winds are observed for the N/C and N/O ratios as seen in Fig.~\ref{NCNOlogg_z02_fin}, we do not include these diagnostics in this Section because determining the baseline CNO abundances in the SMC is not a trivial task.
	By default, our evolution models use the abundances given by \citet{asplund09} rescaled to $Z=0.002$ ($=0.14Z_\odot$).
	However, different studies applying different tracers such as B-type stars \citep{rolleston03,hunter09,dufton20} or supernova remnants \citep{dopita19}, have determined distinct initial values for N/C and N/O ratios, thus making impossible to set a reliable baseline for CNO abundances in the SMC.
	Therefore, we concentrate our analysis and discussion about chemical evolution on the N/H ratio.

	By taking the effective temperatures and surface gravities, we readapt our evolutionary tracks to the spectroscopic HR diagram (sHRD), which is analogous to the original HR diagram but showing a plane of $T_\text{eff}$ vs $\mathcal L(=T_\text{eff}^4/g)$ instead of the luminosity $L_*$ \citep{langer14}.
	The main advantage of the sHRD is that both quantities, effective temperature and surface gravity, are spectroscopically determined and then they are distance-independent, which is particularly important for the study of sources in other galaxies.
	For the analysis of the chemical enrichment we provide Hunter diagrams \citep{hunter09}, which relate nitrogen over hydrogen abundance (by number) to the projected rotational velocity ($\varv\sin i$).
	The selected stars, together with their parameters determined by \citet{bouret21}, are shown in Table~\ref{stars_bouret}.
	
	
	The results for both diagrams are shown in Fig.~\ref{SHRD_z02_fin}.
	Because the evolutionary stage of O-type giants and supergiants correspond either to the late H-core burning stage and the beginning of the He-core burning stage, the helium at surface shown at Table~\ref{table_MS_z02} is comparable with the helium mass fractions shown in Table~\ref{stars_bouret}.
	Therefore, helium mass fraction can help us to expand our comparison between old and new models, especially if we consider the large uncertainties in our analysed stellar sample.
	Hence, we observe that for AV 439, whose high rotational speed suggests that it is still in the MS, shows a better agreement with both the nitrogen enrichment and the surface helium predicted by our new $20\,M_\odot$ evolution model, because its $Y_\text{He}=0.308$ lies below the final value expected at the end of the H-core burning stage.
	Indeed, old $20\,M_\odot$ model would never reach such helium mass fraction, not even beyond the H-core burning stage, as shown in Table~\ref{table_HeC_z02}.
	Likewise, stars AV 307 and AV 43 are also located in the Hertzsprung gap of the old $20\,M_\odot$ evolution model and with good agreement respect to the expected nitrogen enrichment; but their large helium mass fractions ($Y_\text{He}=0.324$ and $0.374$ respectively) match better with the predicted $Y_\text{surf}$ by the new $20\,M_\odot$ model.
	Similar conclusions can be deduced for the other stars, for example: AV 15 ($T_\text{eff}=39.0$ kK, $\log g=3.61$) matches to our $60$ $M_\odot$ (during H-core burning) and $40\,M_\odot$ (end of H-core burning) models, but its poor helium enrichment ($Y_\text{He}=0.285$) suggest that the $60\,M_\odot$ scenario is more realistic.
	Same for AV 95 ($T_\text{eff}=38.0$ kK, $\log g=3.7$), this case matching the MS bands of the models with $40$ and $32$ $M_\odot$ with its $Y_\text{He}=0.285$.
	
	For the stars showing larger nitrogen enrichments, AV 75 ($T_\text{eff}=38.5$ kK, $\log g=3.51$) shows a match with our $85\,M_\odot$, in rule with its MS position in the s-HRD, thus suggesting this star was born with a larger initial mass.
	On the other hand, AV 327 ($T_\text{eff}=30$ kK, $\log g=3.12$) also shows a large nitrogen enrichment but this is not in rule with its position in the s-HRD; however, its helium enrichment ($Y_\text{He}=0.374$) agrees with the predictions of the new models as well.
	The only two stars whose larger nitrogen abundances do not match with any of the new tracks in the Hunter diagram are AV 232 and AV 83, which also are the stars with the largest mass fraction of helium at the surface ($Y_\text{He}=0.444$), thus suggesting that these stars also were fast rotators at the beginning of their lives in agreement with \citet{hillier03}.
	
	In the other extreme of poorly nitrogen enriched stars we have AV 77, AV 47, and AV 69.
	These first two stars are located in the MS band of the sHRD, but only AV 47 shows a helium mass fraction ($Y_\text{He}=0.285$) below the expected at the end of the H-core burning for old and new $25\,M_\odot$ models.
	Because of its age of $\sim6$ Myr, its lower nitrogen abundance might be product of being a slow rotator ($\varv\sin i=60$) in the last part of its MS lifetime.
	For the case of AV 77, evidence of binarity is detected in its spectrum \citep{bouret21}, and then more information about its companion is required to adequately describe the evolutionary status of the star.
	The most extreme case of poorly nitrogen enriched star is AV 69 (even lying out of the plot range of Fig.~\ref{SHRD_z02_fin}).
	The deficient nitrogen abundance of AV 69 is even below the baseline N abundances assumed for the SMC, thus suggesting that this star is a slow rotator with negligible mixing \citep{hillier03}.
	In a minor scale, a slightly lower rotation could explain the abundances of AV 170, which is located in the Hertzsprung gap of our $25\,M_\odot$ model but its $Y_\text{surf}=0.324$ lies marginally below the predicted by new models (but still above the predictions of old models).
	Therefore, our new evolution models can describe the observed properties of O-type stars at metallicity $Z=0.002$, with the natural exception of fast and slow rotators (i.e., both with $\varv_\text{ini}/\varv_\text{rot}\ne0.4$) and binary stars.
	
	In synthesis, evolution models adopting weaker winds for the main sequence not only modify the predicted tracks over the HR and sHR diagrams, but also the chemical diagnostics expected for massive stars specially at the end of the MS.
	Because of this reason, the most optimal observational data to use in our diagnostics corresponds to stars at `evolved' stages such as O-giants and supergiants rather than `unevolved' O-dwarves.
	We can conclude different evolutionary status for the evolved O-type stars from \citet{bouret21} based on our new models, notwithstanding that such conclusions cannot be the final word, because the analysed sample contains only 13 stars and because the observational uncertainties are still significant.
	We expect to be able to draw stronger conclusions in the near future thanks to upcoming projects such as XShootU \citep{vink23a,hawcroft23}, exploring also other spectroscopic phases such as BSG (Bernini-Peron et al. 2024, in prep.) and even RSG if possible.
	
\section{Summary and conclusions}
	This work is an extension of the previous evolution studies from \citet{alex22b,alex23a}, where we upgrade the mass-loss formula for the winds of massive stars adopting the m-CAK prescription \citep{alex19,alex22a} and a lower metallicity.
	Here, we study the impact in a low metallicity environment as the SMC ($Z=0.002$), where the overall mass loss are weaker.
	In concordance, we observe that the most important repercussion of new winds (which go from $\sim2-6$ times weaker than old winds) is not given by the amount of removed mass (as it happens for more massive scenarios at higher metallicities) but for the level of chemical evolution observed at the stellar surface.
	These divergences in chemical evolution are because, weaker winds makes stars lose less angular momentum and therefore keeping a fast rotational speed for a prolonged time, thus producing a more efficient rotational mixing.
	
	For our $20$ to $40\,M_\odot$ models, stars end their H-core burning stage more chemically evolved when new winds are adopted, whereas $60$ and $85\,M_\odot$ models end with less chemically evolved surfaces.
	These results are important for spectroscopical diagnostics, specially those which provide values for He and CNO abundances of massive stars in the SMC.
	However, the accuracy of such diagnostics depends not only on the quality of spectra but also the evolutionary stage of the sources.
	O-dwarves from \citet{bouret13} are too young to use them for comparison between old and new evolution models, whereas O-giants and supergiants from \citet{bouret21} could give us more insight if we exclude fast-rotators and slow-rotators.
	Notwithstanding, analysis on chemical evolution needs to be extended to latter spectroscopic stages such as BSG and RSG.
	Concerning BSG, new values for He and CNO abundances have been obtained thanks to the XShootU project (Bernini-Peron et al. 2024, in prep.).
	In the same line, we are preparing extensions of evolution models covering He-core burning stage for Galactic and LMC metallicities, to compare with the BSG properties found by \citet{bernini-peron23}.
	Concerning RSG, they are important to explore because they reach the largest stellar expansion, which is a key factor for the evolution of close binaries as it affects mass transfer interactions and in turn the formation of GW sources \citep{romagnolo23}.
	Upgrades on the line-driven wind recipes barely modify the $R_\text{max}$ values, thus suggesting that other features such as core overshooting and semiconvection also need to be upgraded \citep{schootemeijer19,martinet21,gilkis21}.
	Not to even mention the most recent studies about the dust-driven winds for RSG \citep{beasor20,kee21,decin24}, to replace the old $\dot M$ prescription from \citet{dejager88}.
	
	The mass-loss formulae adopted for our evolution models from $20$ to $40$ $M_\odot$ are the state-of-the-art, whereas the models for $60$ and $85$ exposed in this work have the purpose of compare the effect of new winds during main sequence but they still need to update mass-loss recipe for later stages.
	The new recipes for WR stars from \citet{sander20a} and \citet{sander20}, together with thick winds in the proximity to the Eddington factor \citep{grafener11,bestenlehner20}, are beyond the scope of this paper.
	
	The results of this work reinforce the importance of upgrading the mass-loss prescriptions in the study of the evolution of massive stars, in particular for their earlier evolutionary stages, even for low metallicity environments where the removed mass during the stellar lifetime is expected to be modest.
	The upgrades introduced in \citet{alex22b,alex23a} and here however, represent the current state-of-the-art related to stellar winds but they are not the final word concerning stellar evolution.
	These studies needs to be complemented by upgrades in other aspects of the stellar structure, such as the convective core overshooting \citep{martinet21,scott21,baraffe23}, the revision of the initial rotational velocities to be adopted, and naturally the upgrade of the mass-loss formula for the advanced stages such as RSGs and WRs.
	These features play a major role for more massive stars, which will be explored with more details in a forthcoming work.
	
\begin{acknowledgements}
	We thank to the anonymous referee for their valuable comments.
	ACGM acknowledges the support from the Polish National Science Centre grant Maestro (2018/30/A/ST9/00050).
	ACGM and JC acknowledge the support from the Max Planck Society through a "Partner Group" grant.
	JC acknowledges financial support from FONDECYT Regular 1211429.
	GM and SE has received funding from the European Research Council (ERC) under the European Union’s Horizon 2020 research and innovation programme (grant agreement Nº 833925, project STAREX).
	SE also acknowledges the support from the Swiss National Science Foundation (SNSF) 
	MC acknowledges the support from Centro de Astrofísica de Valparaíso, Chile, and is also grateful the support from FONDECYT project 1230131.
	We dedicate this paper to Prof. Krzysztof Belczyński, who contributed to this research before his untimely passing on the 13$^\text{th}$ of January 2024.
\end{acknowledgements}

\bibliography{smc_paper.bib} 
\bibliographystyle{aa} 

\end{document}